MAGICSKY– GRANT AGREEMENT NO 665095

# REPORT ON D1.1 (LEAD: PSI)

## Report on imaging of individual skyrmions of MML systems made by different techniques

**Vincent CROS, Roland Wiesendanger, Christopher Marrows, Robert Stamps, Stefan Blügel, Stefan Heinze, Jörg Raabe, Christoforos Moutafis, et al**

31/08/2016

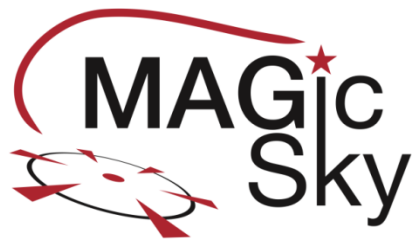

# Abstract

Deliverable 1.1 is a report on imaging of individual skyrmions of MML systems made by different techniques. This deliverable falls within work package 1 (*WP1:Static equilibrium properties of interface induced skyrmions in magnetic films and multilayers*) of the MAGicSky research proposal whose objectives are: the growth of magnetic multilayer systems with large interfacial chiral interaction, the determination of the amplitude of the Dzyaloshinskii–Moriya interactions (DMi), the detailed structural and magnetic characterization of interfaces, the magnetic imaging of individual skyrmions and the thermal stability and interaction of skyrmions with intrinsic and/or artificial defects. It should be emphasized that **imaging magnetic skyrmionic structures forms the basis** for all the planned subsequent material optimization, electrical transport experiments and devices preparation in the MAGicSky research program.

In particular, imaging is paramount as a technique since it enables us to unequivocally demonstrate chiral magnetic skyrmions by directly imaging i) the effects of engineering multilayer samples with large DMis and ii) the skyrmionic structure and properties. In order to get complementary physical insight on the properties of magnetic skyrmions and to corroborate our research results, we utilise a plethora of key state-of-the-art **advanced magnetic imaging techniques**, namely: i) Spin-polarised Scanning Tunnelling Microscopy **(SP-STM)**, Scanning Transmission X-ray Microscopy **(STXM)** Lorentz Transmission Electron Microscopy **(Lorentz TEM)** as well as, Magnetic Force Microscopy **(MFM)**. These imaging experiments can therefore stand as the base for all the other objectives in *WP1* as well as *WP2* that will study *Dynamic properties: nucleation, propagation and detection of skyrmions*. Imaging has been aimed to test and get feedback from t*he theoretical and numerical investigations of skyrmion systems of WP3*. Finally, for *WP4: Towards skyrmion based devices for information storage and processing*, some of the techniques detailed here in D1.1 will enable us to image and test the first devices and their functionality. In short, the MAGicSky consortium research results in D1.1 constitute a major milestone that acts as a stepping-stone to every other experimental effort and we have achieved it. The results relevant to the D1.1 deliverable were presented in 5 publications and several invited talks at international workshop and conferences:

# Project-related publications

# Content

# I/ Imaging of nanoscale skyrmions by spin polarized scanning tunneling microscopy (SP-STM)

**Partners involved : UHAM and CAU**

Spin-polarized scanning tunneling microscopy (SP-STM) combines the high spatial resolution of STM with spin sensitivity by using spin-polarized tips. The magnetic contribution to the signal then depends on the projection of tip and sample magnetization. Consequently, a tip that is magnetized along its axis is sensitive to the out-of-plane sample magnetization and magnetic skyrmions are imaged as axially symmetric objects (see the sketch in Fig. 1a). When a tip with a magnetization parallel to the surface plane is used, magnetic contrast is observed at sample positions with in-plane magnetization components and one can directly prove that all skyrmions have the same rotational sense of their spin structure, being a direct consequence of the chiral Dzyaloshinskii-Moriya interaction (DMi).

The first observation of individual chiral skyrmions was made by UHAM partner for the metallic bilayer system of Pd/Fe on an Ir(111) substrate as shown in Fig. 1b [1]. Analyzing the atomic-scale magnetic signal across individual skyrmions yields a direct measure of the size and shape of these objects, see plot of the polar angle of magnetization versus distance from the skyrmion center in Fig. 1c). To quantify this data, we proposed an analytic solution to the magnetization profile and demonstrate that indeed the experimental data can be modelled by two overlapping tangenshyperbolicus functions. The evolution of the skyrmion size and shape with external magnetic field is governed by the material parameters, which can be determined by a comparison of the experimental data with solutions from the model calculations.

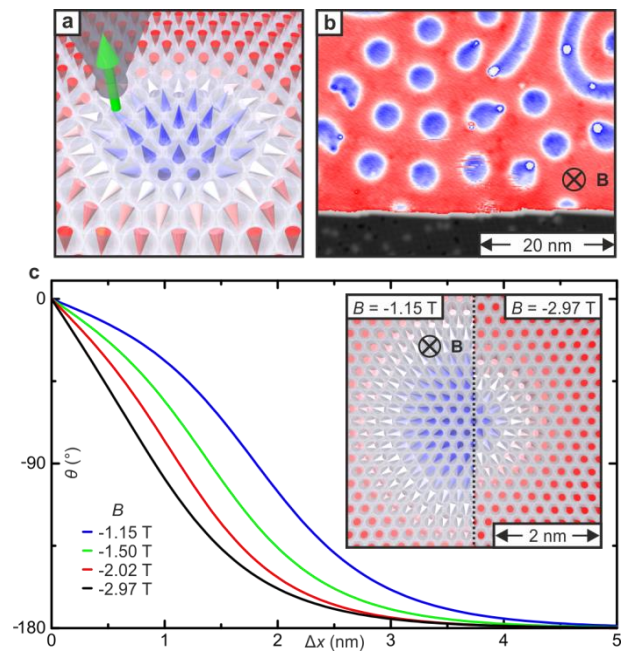

**Figure 1: Skyrmions in PdFe/Ir(111) (a) Sketch of the experimental setup of a spin-polarized tip probing a magnetic skyrmion. (b) SP-STM topographic image of radially symmetric skyrmions in a perpendicular field of $B$ =-1.5 Tesla ($U$=+200 mV, $I$=1 nA, $T$=2.2 K). (c) Theoretical skyrmion size and shape as a function of the external magnetic field. The magnetic parameters, like the DMi, can be determined by a fit to the experimental magnetic field-dependent data.**

The magnetic skyrmions in Pd/Fe/Ir(111) were found to be electronically different to the ferromagnetic surrounding due to the non-collinearity of the spin structure. This discovery, together with a detailed experimental (UHAM partner) and theoretical (CAU partner) study, has led to the introduction of a new magnetoresistive effect, i.e. the non-collinear magnetoresistance (NCMR): in addition to the tunnel magnetoresistance (TMR) which occurs with two magnetic electrodes, and the tunnel anisotropic magnetoresistance (TAMR) which is a material property change due to spin-orbit interaction, we have demonstrated that the NCMR is another important contribution to magnetoresistance in non-collinear magnetic systems. While its large magnitude is comparable to the TMR, it does not rely on the spin-polarization of the probing electrode, thus, the advantages of the two previously known magnetoresistance effects are combined in NCMR. These results have been achieved by a detailed STM study with both spatial (Fig. 2d) and energy (Fig. 2e) resolution in combination with theoretical calculations (CAU).

Subsequent theoretical work from Juelich partner including spin-orbit interaction has demonstrated that both TAMR and NCMR can contribute to an imaging mechanism with non-magnetic tips, and the experimental images are nicely reproduced.

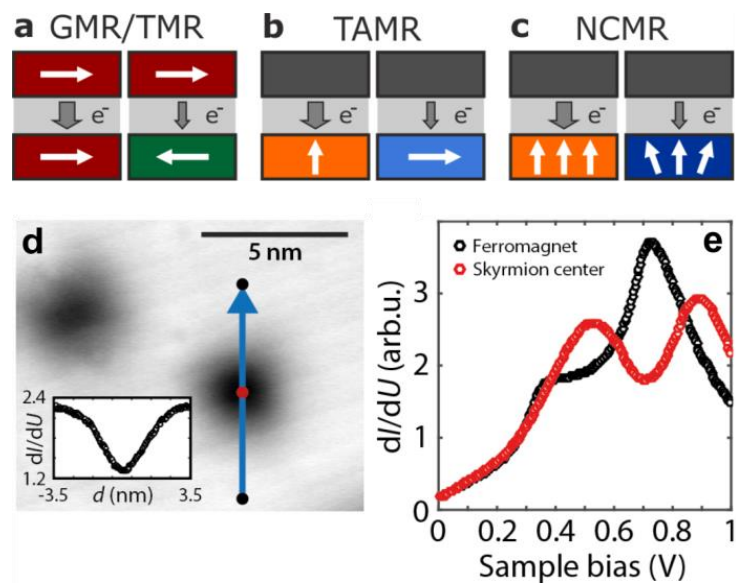

Figure 2: Detecting skyrmions with non-magnetic probe tips via NCMR. Sketches of the (a) tunnel magnetoresistance (b) tunneling anisotropic magnetoresistance and the newly discovered (c) non-collinear magnetoresistance. (d) d$I$/d$U$ map of two skyrmions at $B$=-2.5 T ($U$=700 mV, $I$=1 nA, $T$=4 K) (e) The local magnetic non-collinearity inside the skyrmion leads to drastic changes of the differential conductance, d$I$/d$U$, compared to the collinear ferromagnetic background. Density functional theory (DFT) and tight-binding calculations (CAU) show that the responsible mechanism is the mixing of electronic bands with different spin character.

This novel NCMR imaging mode can be used to image and investigate magnetic skyrmions without the need of spin-polarized probe tips, and it is especially suited to detect spin distortions. Figure 3 shows maps of differential tunnel conductance (d$I$/d$U$) of skyrmions in Pd/Fe/Ir(111) at different canted magnetic field values. Whereas the out-of-plane component in both measurements is +1.3 T, the in-plane magnetic field of 1 T points in opposite directions. Since canted magnetic fields break the axial symmetry of skyrmions their spin structure is distorted, as evident in the two insets with enlarged skyrmions: the signal on the left and right side of the skyrmion changes upon in-plane magnetic field reversal, demonstrating the cycloidal nature of the skyrmions. Comparison with micromagnetic simulations of skyrmions in canted fields enables the determination of the absolute rotational sense of the magnetization, i.e. the sign of the interface-induced DMi [2].

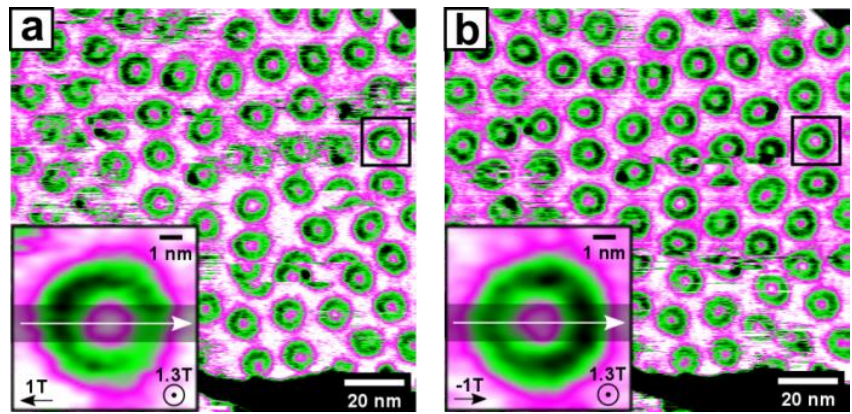

Figure 3 : Skyrmion lattice phase in Pd/Fe/Ir(111), imaged with NCMR (d$I$/d$U$ map, $U$=700 mV). In canted magnetic fields, 1.3 Tesla perpendicular plus (a) 1 Tesla to the left, (b) 1 Tesla to the right, the skyrmions are asymmetrically distorted. This allows to determine the rotational sense (clock-wise rotating), which is in agreement with DFT calculations (CAU).

In general, the size and shape of magnetic skyrmions in Pd/Fe/Ir(111) is determined by the applied perpendicular magnetic field, however, it can deviate from the equilibrium values due to interaction with defects. In Fig. 4a), we show an undistorted magnetic skyrmion in Pd/Fe/Ir(111) at an applied out-of-plane magnetic field of 1 T imaged with NCMR. When Co clusters are adsorbed at the surface, see topographic image in Fig. 4b), we observe significant distortions from the axial symmetric shape due to pinning to these defects and the diameter of the skyrmion is approximately doubled in one direction [3].

Exploiting the strong interaction between magnetic skyrmions and adsorbed Co clusters, we can displace magnetic skyrmions in PdFe by manipulation of a Co cluster with the STM tip, see Fig. 5 [3].

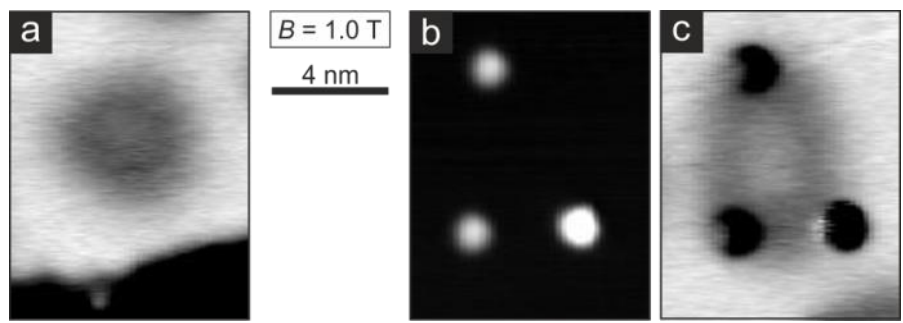

Figure 4: Distorting skyrmions with magnetic clusters. (a) Undistorted skyrmion. (b,c) Skyrmion pinned by three Co clusters at the same magnetic field (d$I$/d$U$ maps, $U$= +700 mV, $I$=1 nA, $T$=4.2 K).

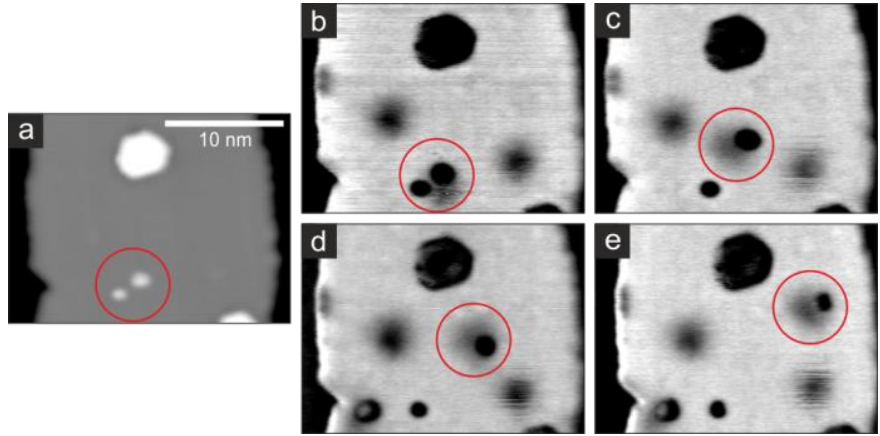

**Figure 5:** Moving skyrmions with magnetic clusters at *B*=2.1 T, *T*=8 K. (a) STM topography (b-e) d*I*/d*U* maps at +700 mV. The skyrmion is pinned to the Co cluster and closely follows when the cluster is moved by the STM tip, see red circle.

A distortion of magnetic states can also be realized by an anisotropic atomic structure of the magnetic material [1]: in a film of three atomic layers of Fe on Ir(111) strain is relieved and reconstruction lines form. This uniaxial structure then gives rise to elongated skyrmions, see Fig. 6 a,b), that reflect the symmetry of the underlying atomic lattice. In a systematic study we have shown that these magnetic objects can be written and deleted in a controlled fashion by electric fields, Fig. 6 a,b) [1,4]. A combination of measurements with out-of-plane and in-plane sensitive tips allows for a reconstruction of the spatially-resolved spin structure, see Fig. 6c) [4].

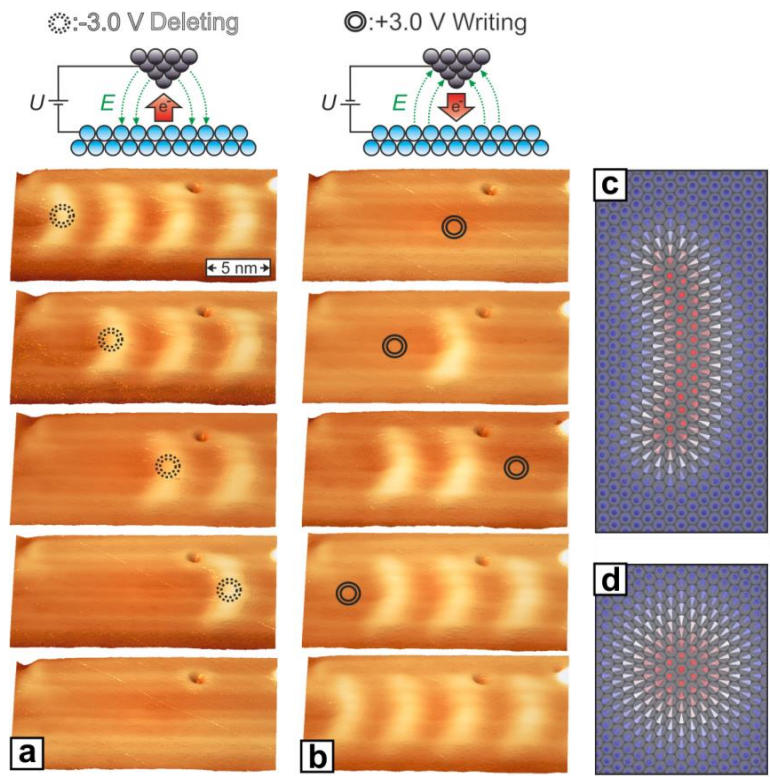

**Figure 6:** Deleting and writing magnetic skyrmions in three Fe layers on Ir(111) with local electric fields. (a) Deleting with local voltage ramps up to -3 V and (b) writing with +3 V. (c) Model of the spin configuration. The skyrmion shape reflects the symmetry of the atomic lattice. (d) Typical skyrmion spin configuration in an isotropic film.

## II/ Room temperature observation of skyrmions by STXM

**Partners involved: PSI, CNRS and UNIVLEEDS**

In this section we focus on static nanoscale magnetic imaging by a X-ray based advanced technique. Specifically, we use Scanning Transmission X-ray Microscopy (STXM) to demonstrate and image for the first time chiral skyrmions at room temperature in technologically relevant multilayers, as well as the evolution of the magnetic skyrmion size with field. In order to achieve this, we engineered tailor-made asymmetric multilayers that exhibit large DMi). We have also initiated the investigations about magnetic imaging of the chiral configuration of skyrmions as well as their thermal stability, that is also part of the research program. The results in this section gather the ones obtained within a core collaboration between the head of the consortium CNRS partner, the Paul Scherrer Institute (PSI partner) and the University of Leeds. The imaging happened on samples grown at CNRS and LEEDS and fabricated at PSI, CNRS or Leeds. The work is complementary and the results form the basis for the subsequent objectives of the MAGicSky research proposal.

**Principle and main characteristics of X-ray magnetic imaging**

All the X-ray imaging experiments were performed using two beamlines at Synchrotron facilities: i) mostly the X07DA (PolLux) beamline at the SLS, Paul Scherrer Institüt, Villigen, Switzerland (PSI partner) and ii) also the Maxymus beamline BESSY II, Adlershof, Germany. As mentioned earlier we utilised the Scanning Transmission X-ray Microscopy (STXM) technique. The magnetic images were recorded by scanning the multilayers grown by sputtering deposition on $Si_3N_4$ membranes with an X-ray beam focused by a Fresnel zone plate, which provided a resolution down to 30 nm. Soft X-rays penetrate the $Si_3N_4$ membrane and the multilayer film. When tuned to an appropriate absorption edge, the X-rays are absorbed by an amount that depends on whether they are right- or left-circularly polarised, and the direction of magnetization in the film: the x-ray magnetic circular dichroism (XMCD). This gives magnetic contrast with, in this case, dark and light areas corresponding to magnetization up or down with respect to the film plane. For some of the images presented here after, the black and white contrast scale has been converted in a post-treatment into a color scale (between red and blue). The X-rays were tuned to the Co $L_3$ edge (779.5 eV) and the measurements were performed at room temperature and under the application of an out-of-plane field. We also developed a set-up for X-ray imaging at low temperatures (down to 100 K) at the PolLux beamline at PSI.

**II.1 CNRS-PSI:** Firstly, we detail the X-ray imaging work from CNRS and PSI which resulted in the demonstration of sub-100nm chiral skyrmions at room temperature (RT) in nanostructures made from technologically relevant multilayers. This was the first observation of such a system internationally [5] and constitutes the completion of a major goal of the MAGicSky research proposal *(Task 1.1)*. The demonstration of room temperature nanoscale magnetic skyrmions and their imaging also stands as a fundamental base upon which the rest of the MAGicSky proposal goals will evolve.

**Nanoscale X-ray imaging of asymmetric magnetic multilayers and estimation of DMi.**

Our approach has been to engineer asymmetric multilayers and to experimentally measure their magnetic domain size or the domain periodicity at zero field after demagnetization. We then compare with micromagnetic simulations (based on experimental material parameters) and determine the size of the DMi needed in order to explain the experimentally observed magnetic domain sizes. We achieve this with independent systematic analyses of the data. In this case, the samples used, which are grown by sputtering, are cobalt (Co) layers interfacing with iridium (Ir) and platinum (Pt) from each side. The reason for the specific choice of the two heavy materials (Pt and Ir) is that they give rise to additive DMi at the two interfaces of the Co layers sandwiched between Ir and Pt. For the results shown here, stacks of ten repetitions of an Ir|Co|Pt trilayer were chosen. Each trilayer was composed of a 0.6-nm-thick Co layer sandwiched between 1 nm of Ir and 1 nm of Pt: Pt10|Co0.6|Pt1|(Ir1|Co0.6|Pt1)10|Pt3 (numbers are thickness in nanometers). Together with these Ir|Co|Pt asymmetric multilayers, we have also prepared symmetric Pt|Co|Pt multilayers as reference samples.

As detailed above, we used (STXM) on specific samples grown on $Si_3N_4$ membranes and measuring the dichroic signal at the Co $L_3$ edge in order to image the out-of-plane component of the magnetization in our Ir|Co|Pt multilayers, as well as its evolution as a function of the external perpendicular field. Selected experimental images measured at different external out-of-plane magnetic field ($H_\perp$) values are shown in Fig. 7 b–e). After saturation at a large negative magnetic field and inversion of the field, we first observed a domain configuration (Fig. 7 b at 8 mT) that combines labyrinthine domains with some domains exhibiting a circular-like shape. When we increase the magnetic field to $\mu_0 H_\perp$ = 38 mT we observe an expansion in size of magnetic domains favored by the magnetic field direction (red domains in Fig. 7 c). Complete saturation of the magnetization is reached in Fig. 7 e) at $\mu_0 H_\perp$ = 83 mT. However, what is very interesting for our purposes is that before saturation, as we see at Fig. 7 d (at 68 mT), there is a typical field range up to about 80 mT in which isolated small magnetic domains of an approximately circular shape persist in an almost totally perpendicularly polarized sample. Since we are using a transmission technique (STXM), we probe the total thickness and therefore these images correspond to an average of the magnetic configuration throughout the ten magnetic layers. Because we mainly observe two opposite contrast amplitudes (corresponding to $m_z = \pm 1$), we can conclude that these magnetic configurations (worms or circular-shaped domains from Fig. 7 b–d)) run throughout the magnetic thickness of the multilayer.

We now focus on the dimension of these circular-shaped domains as well as on their increase when $H_\perp$ is decreased. We demonstrate that they are consistent with a large DMi value and can be modeled as skyrmions. In particular, from the analysis of the STXM images, we can determine precisely the size evolution of selected isolated nanoscale domains at decreasing fields. We find that the diameter of the circular-shaped domains goes from about 30 nm at $\mu_0 H_\perp$ = 73 mT to 80 nm at $\mu_0 H_\perp$ = 12 mT. At a much lower field (closer to zero), the magnetic contrast evolves towards worm-like domains from which a proper diameter can no longer be defined. It is important to note that the diameter of the circular-shaped domains (around 80 nm) that we observe at very low field values remains extremely small compared with the usual values (at the least, larger than 800 nm with the magnetic parameters of our multilayered films) observed in classic bubble systems in which the driving mechanism for bubble stabilization is the dipolar interaction.

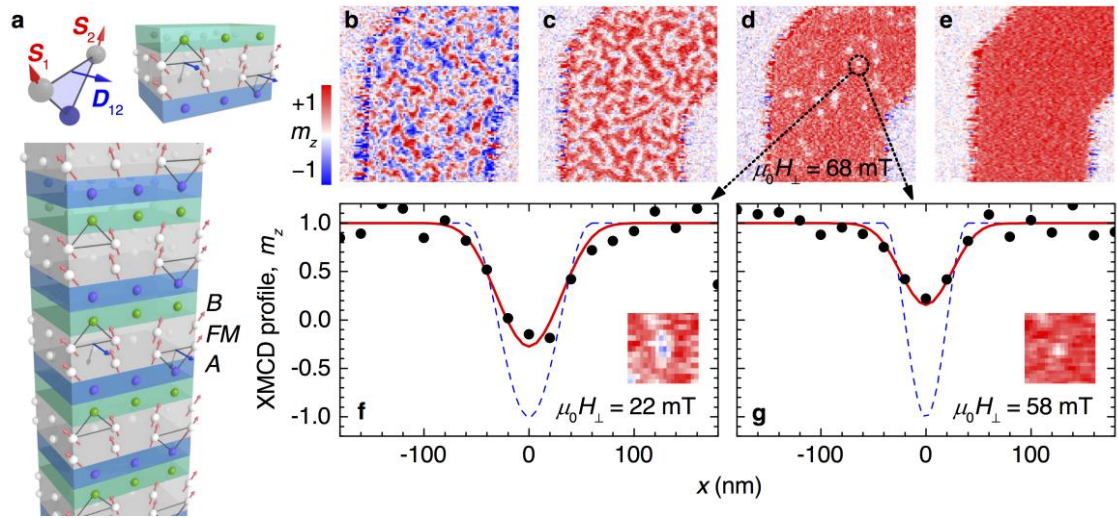

**Figure 7:** Interfacial DMi in asymmetric magnetic multilayers. a, Top left: the DMi for two magnetic atoms (grey spheres) close to an atom with a large spin–orbit coupling (blue sphere) in the Fert–Levy picture. Top right: zoom-in on a single trilayer. It is composed of a magnetic layer (FM, grey) between two different heavy metals A (blue) and B (green) inducing the same chirality (same orientation of D) when A is below and B above the magnetic layer. The bottom shows a zoom-in of an asymmetric multilayer made of several repetitions of the trilayer. b–e), A 1.5 × 1.5 µm2 out-of-plane magnetization ($m_z$) map obtained by STXM on a (Ir|Co|Pt)x10 multilayer at r.t. for in-situ out-of-plane magnetic fields of 8 (b), 38 (c), 68 (d) and 83 (e) mT. The actual image size of the insets is 360 nm. In f) we see the X-ray magnetic circular dichroism (XMCD) signal through a magnetic circular domain (skyrmion) observed at 22 mT (black dots). The blue dashed curve is the magnetization profile of an ideal 60-nm-diameter skyrmion and the red curve derives from the model described in the text. G) We show the same at 58 mT and the corresponding simulation of a 40-nm-diameter skyrmion. The data from f) and g) come from the same skyrmion evolution under the magnetic field.

By comparing the experimental data for the field dependence of the size of the circular-shaped domains with micromagnetic simulations of DMI-induced skyrmions, we determine a very large DMi value of $|D|$ = 1.9 ± 0.2 mJ.m$^{-2}$ (for an exchange constant A = 10pj.m$^{-1}$). In addition, micromagnetic simulations show that already for $|D|$ > 1 mJ.m$^{-2}$, the only possible circular configurations are isolated magnetic skyrmions with a winding number of unity. Therefore, we can determine with certainty that the bubble-like domains we observe are in fact isolated magnetic chiral skyrmions at room temperature. It is important to also note that when we analysed the STXM images of the symmetric (Pt|Co|Pt) multilayers, we did not find such a large DMi like in the case of our engineered asymmetric multilayers (the DMi in the symmetric multilayers is small $|D|$ = 0.2 ± 0.2 mJ.m$^{-2}$, but not negligible, attributed to the different structures of the Pt interfaces with Co).  We also developed a second method that relies on the quantitative analysis of the mean width of the perpendicular magnetic domains from the STXM images at remanence and again compared those with micromagnetic simulations.  After we also take into account the proximity-induced magnetization of Ir and Pt, we find a similarly large DMi of about 1.4 ± 0.2 mJ.m$^{-2}$ < $|D|$ < 1.6 ± 0.2 mJ.m$^{-2}$. It should be emphasised that the domain width we find in Pt|Co|Ir lies in the sub-100 nm range at zero field for which the DMi is the main stabilization mechanism, and, again, cannot be expected with only dipolar interactions, as it would be the case for classical hard magnetic bubbles.

**Room temperature X-ray imaging of skyrmions in nanostructures of asymmetric multilayers.**

As described in the previous section, we found from two independent analyses of the magnetic configurations that very large DMi and skyrmions exist at RT in asymmetric (Ir|Co|Pt)x10 multilayered films. Here we show that isolated nanoscale skyrmions can be stabilised in confined geometries, which is a pre-requisite for technological applications and the MAGicSky objectives in *WP2* and *WP4*. Specifically we demonstrate that individual magnetic skyrmions can be stabilized at room temperature in nanodisks and nanostrips which we pattern from our multilayers by electron-beam lithography and ion-beam etching. Figure 8a) shows the field dependence of the diameter of an approximately circular domain located close to the centre of a 500-nm-diameter disk, and this dependence is compared, as we did in the first section, with that obtained in micromagnetic simulations. Again, a major outcome of these simulations is that it is not possible to stabilize any bubble-like domain in submicro- metre-sized disks without introducing large DMi values of at least |D|=1.5 mJ m–2. Since in our simulations the winding number of the circular domains is, after stabilization, always equal to one, we can conclude that the experimental images correspond to nanoscale skyrmions with a chirality fixed by the sign of D. In Fig. 8b we show X-ray magnetic images of 300 nm disks and tracks ("racetracks") 200 nm wide with even smaller skyrmions that are stable down to very small fields (∼ 8 mT) of dimensions that range from 90 nm close to a zero field to 50 nm in an applied field. We therefore unambiguously show the existence of sub-100 nm magnetic skyrmions in nanostructures and at room temperature.

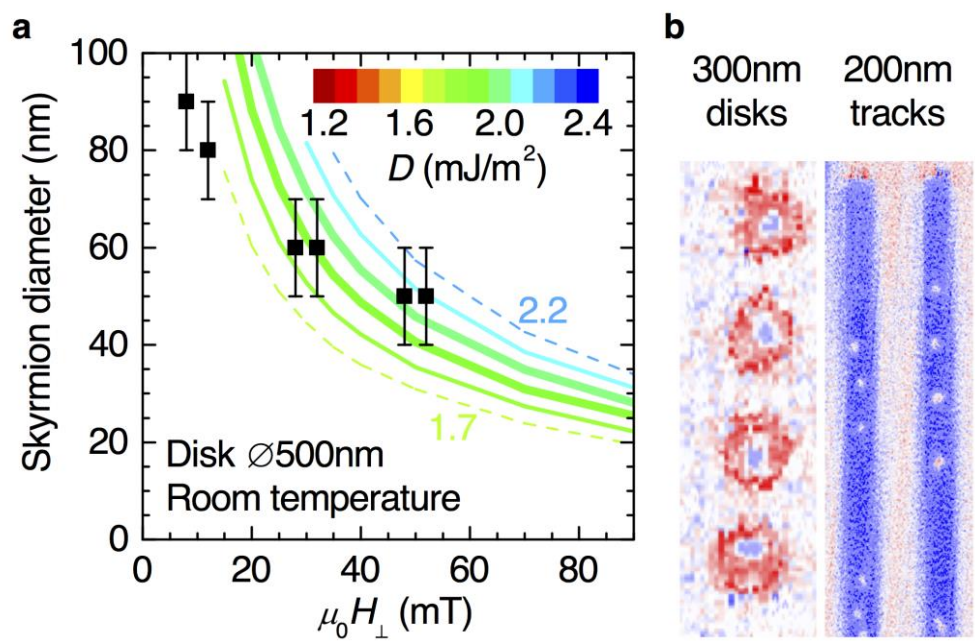

Figure 8: Magnetic skyrmions in patterned nanoscale disks and tracks. a) Magnetic field evolution of the skyrmion size derived from micromagnetic simulations realized for A = 10 pJ.m$^{-1}$ (lines for different |D| values) and the experimentally determined sizes of the observed skyrmions (squares) for 500-nm-diameter disks. B) Left: Room temperature (R.T.) out-of-plane magnetization map of 300-nm- diameter disks at a bias out-of-plane external field of $\mu_0 H_\perp$ = 8 mT. Right: 200-nm-wide nanowires ("racetracks") at $\mu_0 H_\perp$ ≈ 55 mT that display several isolated skyrmions.

**Thermal stability investigation of magnetic skyrmions: Low temperature X-ray imaging of skyrmions in nanostructures of asymmetric multilayers.**

Demonstrating chiral skyrmions at room temperature was crucial goal because it is a prerequisite for devices. One of the next MAGicSky proposal goals, going beyond the scope of deliverable D1.1, is to also investigate their thermal stability. We show here that have already initiated the thermal stability investigation of magnetic skyrmions by nanoscale X-ray imaging at variable temperatures. In order to achieve this new variable temperature set-up was implemented at the PolLux STXM beamline at PSI that enabled us to use X-ray image from 100 K temperatures up to 300 K temperatures.

In Figure 9 a) we see the image of a curved magnetic nanowire along with the adjacent magnetic nucleation pads. In Fig. 9 b) we image, at room temperature, a 6x6 µm area of the nanowire at 80 mT bias perpendicular field where magnetic skyrmions are nucleated/present. In the next step we decrease the temperature down to 100 K and take an image of the same are again. What we see is that in c) (red square) the same skyrmion state is present. It should be noted that this is in contrast to when we start with a saturated magnetic state at room temperature and then lower the temperature. It is then difficult to nucleate skyrmions. However, if we start with a skyrmion state at 300K, then we effectively "freeze" this state down to 100K.

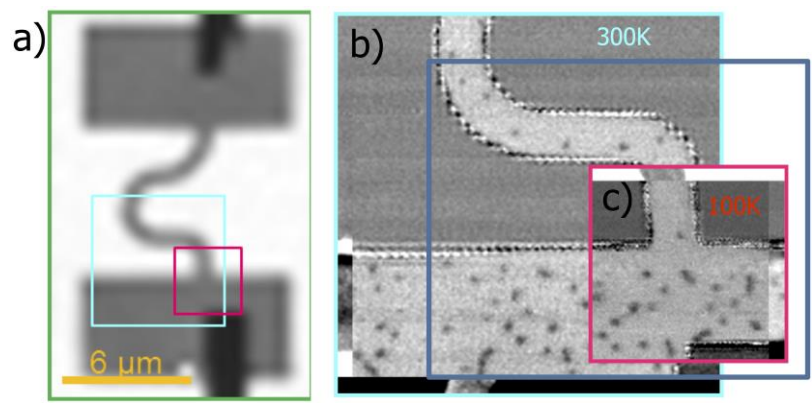

Figure 9: Thermal stability investigation of skyrmions by X-ray imaging at variable sample temperatures. a) Curved nanowire and adjacent nucleation pads made from ||Pt10|(Ir1|Co0,8|Pt1)x20|Pt8 (thickness in nanometers). B) A 6x6 µm magnetic image at 80 bias perpendicular field where a skyrmion state has been nucleated at room temperature (300K). In c) we image the part of the wire indicated in the red square at a very low temperature (100K). We observe the same skyrmion state. We are able to "freeze" the skyrmion temperature from room temperature down to low temperature.

**Nanoscale time-resolved X-ray imaging of magnetic skyrmions**

The first major milestone achieved within the MAGicSky project was to stabilize and image room temperature magnetic skyrmions in nanostructure and engineer a multilayer system with large DMi, as detailed in the above section. The next step, which goes beyond the scope of Deliverable 1.1, will be to achieve control over the motion skyrmions. The best way to achieve that is to directly image skyrmion dynamics and this is a research effort that has already started in the skyrmion community in general. In order to achieve this nanoscale sub-nanosecond dynamical X-ray imaging is required. Therefore, we have already started experiments to dynamically image the behaviour of skyrmions in nanostructures and present here some preliminary results. We use X-ray imaging at the MAXYMUS beamline (STXM) and operate it in a stroboscopic (pump-probe) scheme. We image dynamically with 100 picosecond resolution. As seen in Figure 10a), we use a microcoil where we pass current pulses that result in perpendicular tipping fields as excitations. We image at a bias in-situ out-of-plane magnetic field of 90 mT which results in a multiskyrmion state in a disc-shaped dot adjacent to the microcoil. In Fig. 10b) we see a screenshot from the time-resolved acquisition. We pass a 6 ns pulse where we can strong magnetic contrast but no clear motion is observed. In Fig. 10c) we use a 20 ns pulse and we see in all the frames that the magnetic contrast is lost due to resistive heating from the stripline (microcoil). Our detection scheme results in full dynamical movies. In order to excite considerable dynamics that can be detected, we need to use excitations between the two extremes in Fig. 10b) and Fig. 10c).

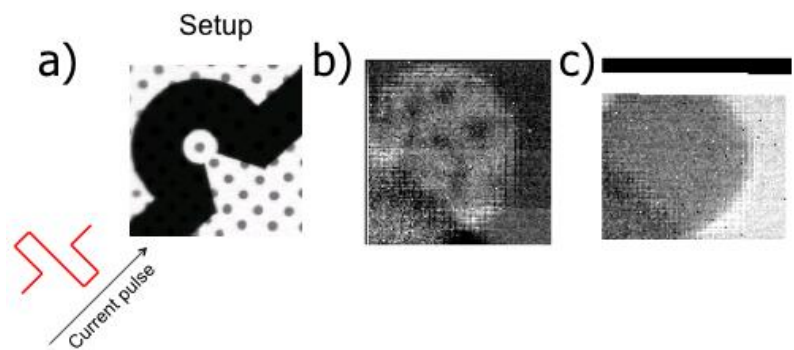

**Figure 10:** a) Patterned disc-shaped dots and adjacent microcoil. The patterned dots are made of ||Pt10|(Ir1|Co0.6|Pt1)x20|Pt3. The resulting field pulses are perpendicular to the sample plane. b) and c) Scanning transmission X-ray microscopy images, acquired at the Co $L_3$ edge, with a temporal resolution of 100 picoseconds. We see here the initial frame. b) A disc-shaped adjacent to the micro-coil is imaged. It exhibits a multi-skyrmion state. The pulsing scheme excitation uses a 6ns pulse. In c) the excitation uses a longer, 20 ns, pulse and has resulted in loss of magnetic contrast due to heating.

**II.2 UnivLeeds-PSI:** In this paragraph, we will focus on static imaging by STXM of chiral spin textures, including worm domains and individual skyrmion bubble domains at room temperature and the effect of applied field on skyrmion size. We show here imaging data of skyrmions at remanence (zero magnetic field) and at room temperature.

**X-ray imaging of skyrmions in nanostructures at remanence and at room temperature**

The asymmetric multilayers grown in Leeds chosen for this experiment were, similarly to before, Pt/Co/Ir thin films and were deposited by sputtering onto $Si_3N_4$ membranes. The films consist of a 4,6 nm Ta buffer layer, a 7.5 nm Pt layer, and then a trilayer of Co(0.4 nm)/Ir(0.5 nm)/Pt(2.3 nm) that is repeated 10 times, finishing with a 1.3 nm Ir layer. The sandwiching of the Co between Pt and Ir, and the 10-fold repeat of the trilayer, results in a strong interfacial chiral interaction that stabilises chiral domain walls and skyrmion bubbles.

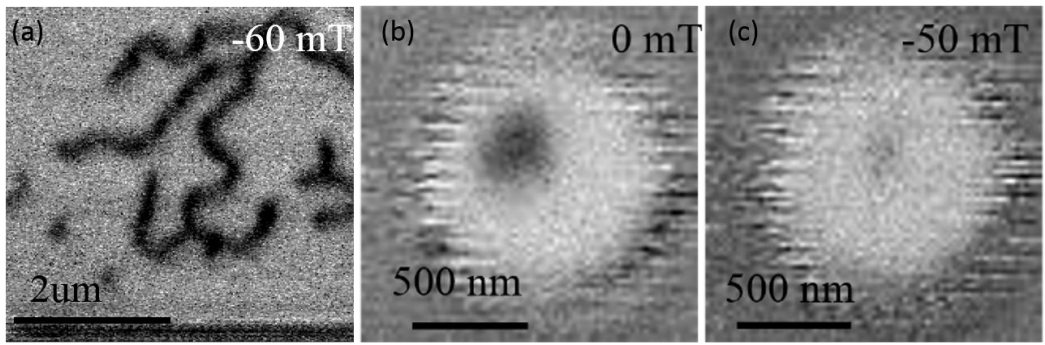

**Figure 11:** Scanning transmission X-ray microscopy images, acquired at the Co $L_3$ edge, of a Ta(46 Å)/Pt(75 Å)/[Co(4 Å)/Ir(5 Å)/Pt(23 Å)]$_{\times 10}$ Ir(13 Å) (a) thin film, and a 1000 nm diameter nanodisc of the same film at (b) 0 mT and (c) -50 mT. Fields are applied perpendicular to the sample plane.

Evidence of this domain structure is seen in Fig. 11, which shows images with dark and lights areas of magnetic contrast from the film and from circular micrometer-sized dots patterned from the film. In the extended thin film, that is not patterned and there are no effects of confinement, shown in Fig. 11a), worm domains form in the majority of the imaged area, and a few bubbles pinch off at sufficiently large applied fields. The bubbles are typically 100-200 nm in diameter. When the film is patterned and confined to micrometer-sized dots, skyrmion bubbles are readily observed at zero applied field, as shown in Fig. 11 b). The bubbles expand and shrink as the applied field is decreased or increased respectively, as in Fig. 11 c).

To conclude this section, we are able to demonstrate nanoscale (sub-100nm) magnetic skyrmions at room temperature and at remanence (zero magnetic field), which are prerequisites for technological devices. In order to achieve this, we engineered multilayers with a large DMi. In view of our future objectives in i) thermal stability, we demonstrated "freezing" of a skyrmion state from room temperature down to 100 K and ii) skyrmion dynamics, we started dynamical X-ray imaging experiments.

## III/ Observation by Lorentz TEM

### Partners involved: UNIVERSITY OF GLASGOW, CNRS

Imaging at Glasgow partner (UNIVERSITY OF GLASGOW) was undertaken on samples fabricated at CNRS. The imaging was performed via transmission electron microscopy in conventional transmission and scanning modes (TEM and STEM) on a JEOL ARM 200cF. Magnetic (Lorentz) imaging was carried on plan view samples prepared on transparent silicon nitride amorphous window membranes which allowed transmission of 200 kV electrons. Additionally, imaging of the interfaces using high resolution imaging and electron energy loss spectroscopy (EELS) were made on cross-section samples prepared using focused ion beam (FIB) "lift out" methods. The samples imaged were multilayer structures with a repeat unit of Ir/Co/Pt for large strength DMi, two such samples grown specifically on ultra-thin membranes were investigated (layer thickness in nm) : *Pt(3){[Ir(1)/Co(0.6)/Pt(1)]×N}Pt(10)where N = 10, 20*

The physical and compositional structures of the samples were undertaken by making TEM cross-sections using the FIB. Such sample allow the interfaces to be imaged as the electron beam is incident parallel to the interfacial planes. For best imaging the thickness of the cross-section should if possible be no more than 50 nm. The sample produced were slightly thicker than this but were thin enough to give a good characterization of the layer structure. Figure 12 shows the results of high resolution imaging and EELS on the cross-section with 10x repeat structure. Fig. 12a) is a high angle annular dark field image (HAADF) of the section which is sensitive to atomic number i.e. elements with a higher Z appear brighter. The 10 dark Co layers are shown clearly in the centre of the image, indicating well-defined and continuous layers clearly separated by the lighter contrast Ir and Pt layers. The green area highlighted in Fig. 12a) was then investigated with EELS and processing of the EELS data provided elemental maps of the key components of the structure. These are shown in Fig. 12b) for Pt, Ir, Co, Cu and N. This nicely shows the definition of the key active layers of Pt, Co and Ir. The Cu was a protective cap layer put on the surface before the normal protective Pt, both administered in the FIB. The N locates the silicon nitride substrate. Furthermore, the data can be plotted as elemental profiles and this is displayed in Fig. 12c) for the active elements in the stack. Again the quality of the layer definition is evidently demonstrated here. The pixel spacing of 0.5 nm is sufficient to sample multilayer structure with layers of 0.6 and 1.0 nm. It should be noted that the percentage level shown on the vertical axis suggests that the layers have much less than 100% of the elements in each of the active layers, apart from the thicker (wider in this case) Pt layers. The reason for this is the thickness of the cross-section (60-70 nm) which means that the electron beam samples the adjacent layers for the thinnest (narrowest) layers here. Thinner cross-sections will give better quantification of the percentages however the data here gives very powerful evidence of the high quality layer structure and definition.

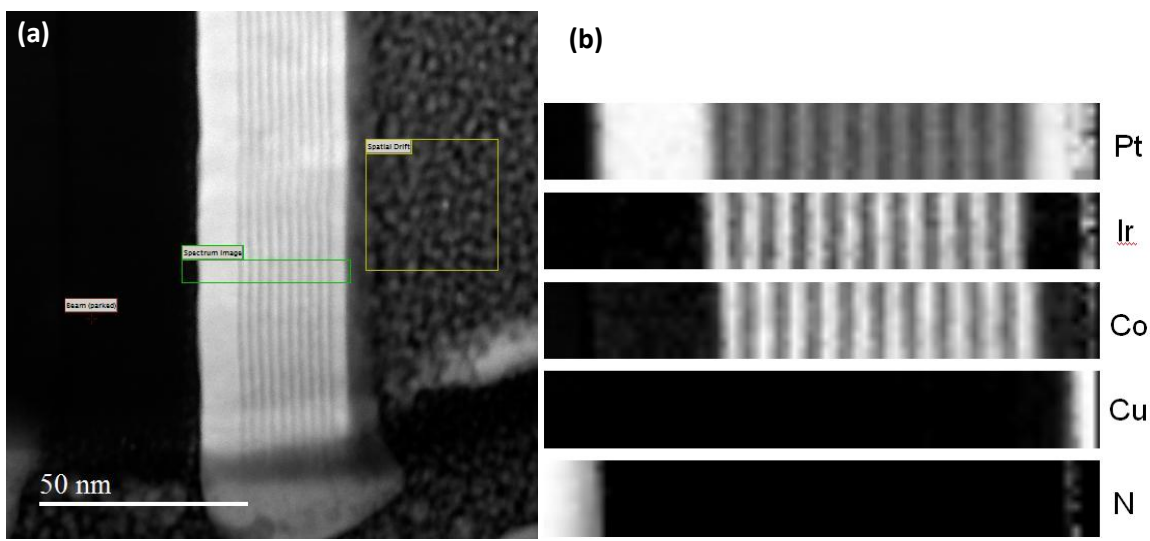

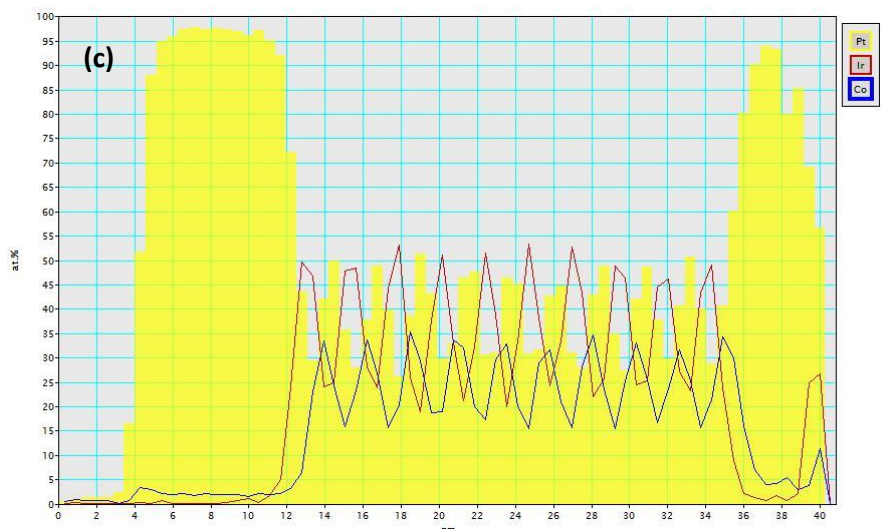

Figure 12: (a) HAADF image of cross-section of 10x multilayer structure showing layers and area (green) for EELS analysis. (b) Elemental distribution in region defined in (a) highlighting well defined active Pt, Ir and Co layers plus substrate (N) and protective layer (Cu). (c) Linetraces of variation of active elements in ML stack.

Magnetic imaging was performed in the same microscope, however in order to image the magnetic structure, the objective lens must be switched off. The field from this lens when fully excited is around 2.0 T which is more than enough to fully saturate the magnetization of samples. Imaging here was performed in TEM with the Fresnel defocused mode in operation. For sample with in-plane magnetization, the Lorentz deflection of the electron beam either side of a domain wall will cause the electron to converge or diverge. By defocusing the walls then appear as bright and dark lines depending on the sign of defocus. In the case of the samples imaged here they possess out of plane (perpendicular) anisotropy. Therefore, at normal incidence, the domains will provide no deflection of the beam and hence no contrast. However, in the region of the domain wall the magnetization will have an in-plane component. Two possibilities exist for a simple one dimensional wall, being either Bloch or Néel-type. To illustrate this, the structure of these walls are shown for a circular domain wall in Fig. 13 a) and b) for Bloch and Néel wall respectively. The Bloch wall is divergent free and the in-plane magnetization (induction) within the wall will cause a divergent/convergent effect of the electron beam at either side resulting in black/white contrast at normal incidence of the beam. This is shown schematically in Fig. 13 c) where black and white contrast is shown either side of the wall on a grey background. In comparison the Néel wall has in-plane magnetization, but in this case this component of magnetization is has divergence and the resulting stray field cancels and deflection to the magnetization. The wall is therefore invisible at normal incidence. However, on tilting the sample, the magnetization from the domains then produces a component of induction perpendicular to the electron beam resulting in domain wall contrast as shown schematically in Fig. 13 d). It should be noted that the wall contrast is seen where the magnetization either side of the wall lies parallel to the wall, whereas where the magnetization points head to head or tail to tail no contrast is visible. The latter is gain due to the divergence of the magnetization. Thus whilst the Néel wall becomes visible there is no continuous contrast along the wall, the discontinuity indicates the tile axis. The main point to note here is that by analyzing the structure of the domain walls and their effect on the electron beam it is possible to distinguish between Bloch and Néel walls. Tilting of the sample with Bloch walls would show a superposition of the contrast seen in Figs 13 c) and d). Bloch walls are expected for materials with perpendicular magnetic anisotropy due to magnetostatic considerations, whilst the presence of an interfacial Dzyaloshinskii-Moriya exchange interaction will favour Néel walls.

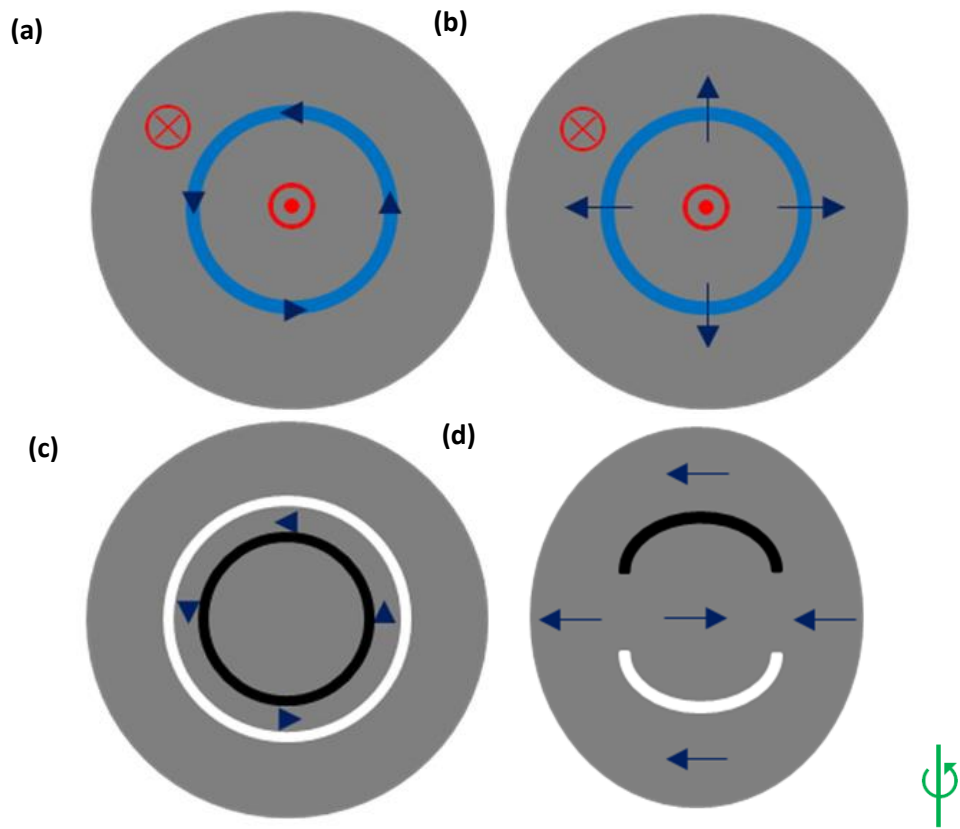

**Figure 13: Schematic of (a) Bloch and (b) Néel wall in thin film with perpendicular magnetic anisotropy. Schematic of Fresnel images of (c) Bloch wall in untilted film and (d) Néel wall in tilted film. The tilt axis is shown by the green line and arrow.**

When imaging these multilayers in the TEM, it is necessary to be aware of some of the challenges faced due to the fact that the majority of the magnetization lies out of the film plane. As already stated at normal incidence of the electron beam this component of magnetization gives not contribution to magnetic contrast. When tilted away from this orientation by an angle θ, the component of magnetization which gives a Lorentz deflection becomes Msinθ. For the TEM membrane samples the tilt is restricted to around 25$^o$ due to the etching of the silicon support. This means that with the small thickness of sample the resulting Lorentz deflection angles from the domain will be very small. As a guide the 20x layer would have a deflection of less than 1 µrad at a tilt of 20$^o$ compared to an equivalent in-plane magnetized sample 12 nm thick Co which would have a deflection about an order of magnitude larger. A further complication is that the electrons have to traverse the whole stack which for the 20x is 65 nm and effectively more when tilted. This increases inelastic scattering of the electrons and will degrade the magnetic contrast.

The imaging of the films was all made with the Fresnel mode of Lorentz TEM. Firstly, images were taken with the sample untilted and tilted to determine the wall type. An example of an untilted and tilted image is shown in Fig. 14 a) and b) respectively. Both images are defocused with a defocus distance of 10 mm. It can be seen from Fig. 14 a) that there is little in the way of black/white contrast, the main contrast is a darkening of the image in the upper left corner due to some bending resulting from stress experienced by the membrane. Imaging of the same region but tilted by 25° shows distinctive small scale black/white contrast which is eliminated by applying a field. The latter indicates the contrast is indeed magnetic, and the non-visible magnetic contrast untilted is consistent with the walls being Néel in character.

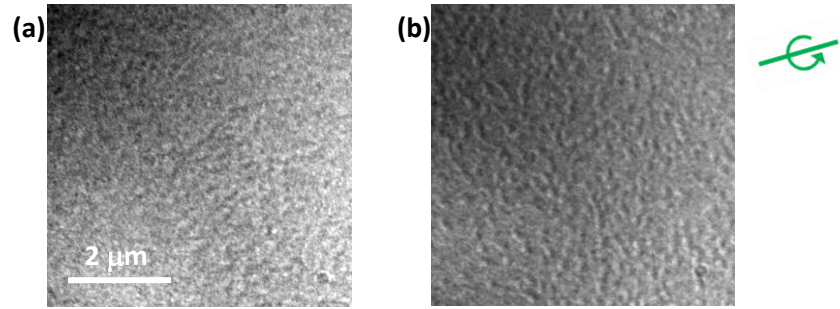

Figure 14 (a) Untilted Fresnel image from 10x multilayer structure showing no magnetic contrast. (b) Fresnel image from same area as (a) but tilted by 25° in the direction indicated by the by the green axis and arrow, showing magnetic contrast.

Additionally, Fresnel imaging can be used to characterize the domain sizes and therefore deduce the magnitude of the DMi constant D. This is done by looking at the remanent state of the films where there is equal up/down magnetised regions. As example of this is shown in Fig. 15a) which is a Fresnel image taken from the 10x layer sample. The image is analysed by taking a fast Fourier transform, which is shown in Fig. 15b). As the image is defocused there are a series of rings visible which effectively show the contrast transfer function of the microscope. Also visible are two arcs which relate to the spatial frequency associated with the skyrmion domains. From these arcs the domain size can be measured and for the two ML samples here the domain size was measured as 125 nm (20x) and 120 nm (10x), effectively identical within error. By analysing the domain size for these samples based the periodicity of the domains can be used to estimate the domain wall energy and hence the DM exchange constant D. Using material parameters which can be measured $M_s = 8\times10^5$ A/m, $K = 10^6$ J/m$^3$ and the ferromagnetic exchange constant A (for such a system quoted in range $1-1.5\times10^{-11}$ J/m) provides and estimate for D in the range 1.6-2.3 mJ/m$^2$ for a domain periodicity of 240 nm, that is in agreement with estimation presented before by STXM analysis on the very same multilayers.

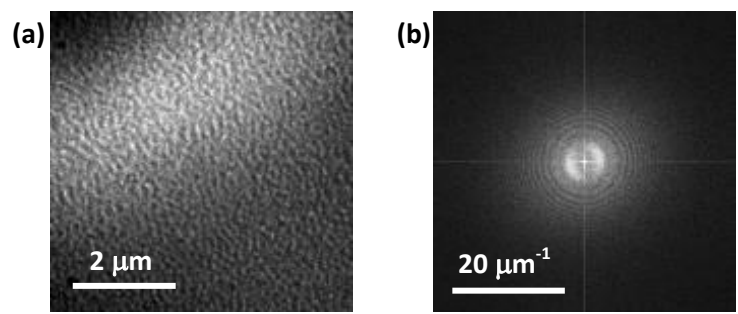

**Figure 15 (a) Tilted Fresnel image from 10x multilayer structure showing magnetic contrast. (b) Fast Fourier Transform of (a) showing distinct spatial frequency associated with magnetic structure allowing estimate of domain size.**

The images in Figs. 14 and 15 highlight the difficulties in terms of low magnetic contrast from the PMA multilayered samples and the overall thickness of material presented to the electron beam. Whilst the average domain size can be extracted as indicated in Fig 15, imaging of individual skyrmions is more challenging. Taking the 20x ML film through an in-situ field sequence in the TEM does show that, for these films the skyrmions form locally and are quite stable. In Fig. 16, the 20x layer is shown at a number of different field values from a uniformly magnetized state (a) and then with decreasing field (b)-(e). No domain structure or skyrmions are seen in (a), but when they appear as can be seen in (b)-(e) it is evident that they have the dimensions measured in Fig. 16, i.e. individual skymions are around 120 nm in diameter and are stable in the applied field.

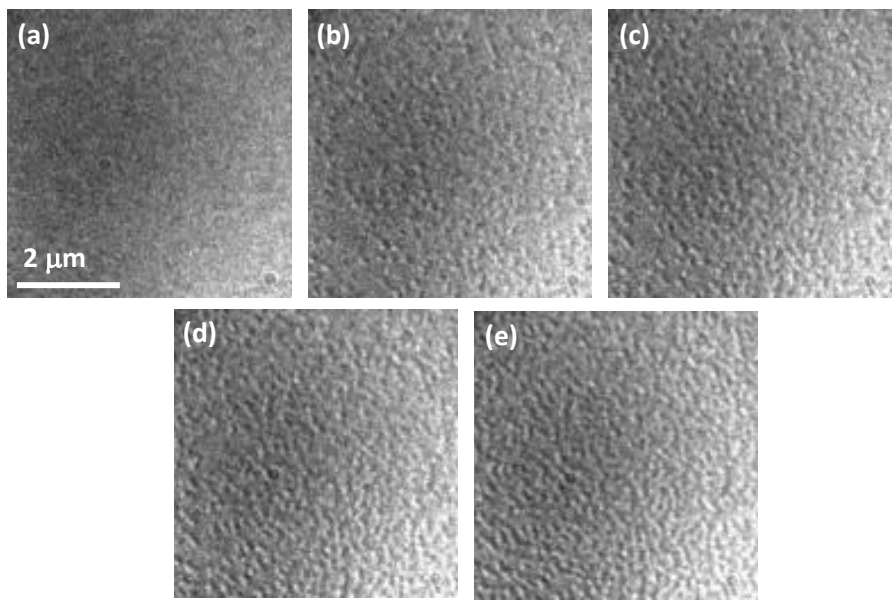

**Figure 16. Tilted Fresnel images from 20x multilayer structure showing evolution of domain structure. The objective lens field applied for the different images here is (a) 1040 Oe, (b) 709 Oe, (c) 626 Oe, (d) 545 Oe and (e) 463 Oe.**

In summary, at this stage, we can see that TEM has shown the asymmetric multilayers which give rise to DM interactions are extremely well defined from the EELS and high resolution study of the cross-section which show continuous layers even for the thinnest layers which are under 1 nm thick. Lorentz microscopy has been used to prove that the skyrmions/domain walls observed in these films are of a Néel character with a domain size periodicity of 120/240 nm measured using FFT analysis of the images. This provides an estimate of the DMi constant D for this film stack of in the range 1.6-2.3 mJ/m$^2$, depending on the value of exchange constant used. The skyrmion structure has been shown to be stable in the presence on an applied field as viewed in-situ in the TEM.

# IV/ Room temperature imaging of skyrmions by MFM (CNRS)

**Partner involved: CNRS**

In this section of the deliverable report, we aim to present the detailed characterization by Magnetic Force Microscopy (MFM) of the magnetic configuration obtained at room temperature in asymmetric multilayers grown by sputtering deposition by the CNRS partner. This imaging technique that is accessible at the lab has to be seen as a complementary imaging technique with the observation first made by STXM and presented in Section II above.

Here we report room temperature imaging on different systems i.e. Pt/Co/Ir, Pt/Co/W, Pt/Co/AlOx by using Magnetic Force Microscopy. Importantly, we demonstrate imaging of magnetic skyrmions confined in disk-shaped nanostructures at a zero bias magnetic field. We also show the dependence of skyrmion properties on the number of repetitions of the multilayers.

## IV.1 Principle and main characteristics of magnetic imaging by Magnetic Force Microscopy

The Magnetic Force Microscopy (MFM) is a variant of the Atomic Force Microscopy (AFM). It records the magnetostatic forces or force gradients between a magnetic sample and a tip covered with a thin ferromagnetic film. This technique has been extremely used and useful in the field of room temperature imaging magnetic patterns in perpendicularly magnetized media, similar to the ones we aim to study in MAGicSky and in which magnetic skyrmions can be stabilized at room temperature. The two most prominent advantages of MFM contributing to its success are its potential insensitivity to non-magnetic surface coatings and reliefs, and good resolution down to about 30 nanometers. The magnetic transition geometry and stray field configuration in perpendicular recording media is illustrated in Fig. 17. The MFM set-up that has been used includes a module allowing to sweep the applied magnetic field in the perpendicular direction. The maximum field amplitude is about ± 350 mT.

Following the same strategy to estimate the amplitude of the DM interaction as the one we introduced from the analysis of the STXM images, we have used the MFM technique to investigate the alteration of DMi by changing several parameters of the multilayers, namely the number of repetitions or thickness of the magnetic layer. The results of this thorough investigation will be presented later. However, it is to be noticed that with MFM, the contrast is not related to the sample magnetization as it is the case in STXM experiments but to the stray field from the magnetic multilayers. Therefore, in order to extract quantitative information from MFM images, we have compared the experimental images with the ones obtained with the micromagnetic solver Mumax3 which has built-in generation of MFM images from a 2D magnetization. The MFM tip lift can be freely chosen. By default, the tip magnetization is modeled as a point monopole at the apex. This means that, at the end of the micromagnetic simulations, we are no longer only recording the map of magnetization, but we are also able to generate a simulated MFM image corresponding to the MFM lift height used as inputs in the parameters. The general lift height chosen for imaging all the samples is 20 nm, similar to the one used in real experiments.

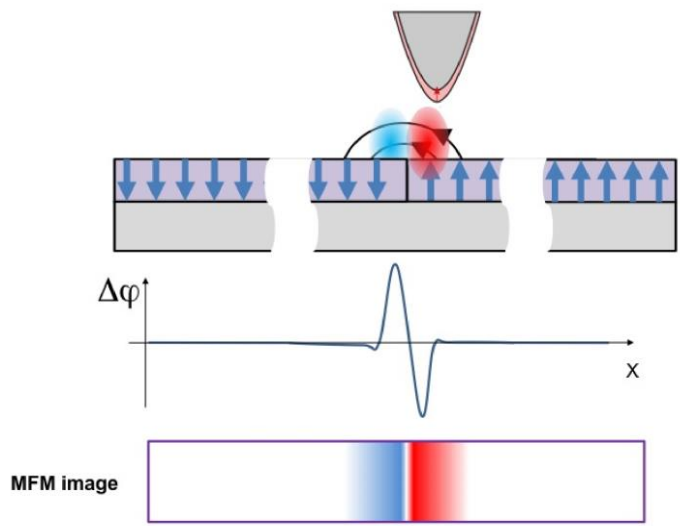

Figure 17: (top) : the magnetic tip of the MFM is scanning a sample with several perpendicular magnetic domains. (Middle), the shift in the phase induced by the interaction between the tip and the stray-field of the sample. (bottom) : Magnetic contrast obtained with MFM.

Indeed, in order to have a better control on the error margin during the experiments, we have performed a preliminary study of the variation of the apparent size of skyrmions as well as the periodicity of wormy (maze) domains versus the height of the MFM lift of the tip. It is also to be pointed out that a second check of the validity of this approach is that in the case of the worm domain, the Fourier transformation for the actual magnetization map and for the MFM simulated image give the same result for the domain periodicity. Thus, in the next paragraphs, to estimate the DMi, we present the comparison of the size of individuals skyrmions as seen by MFM with the one simulated with MFM tip lift of 20nm. All the other input parameters i.e. saturation magnetization and magnetic anisotropy have been determined experimentally through magnetometry measurements. Finally, the exchange constant is chosen to be 10 pJ/m (see C. Moreau-Luchaire *et al.*, Nanotechnology 11, 444–448 (2016)).

### IV.2 Systematic MFM studies and estimation of the amplitude and sign of DM interaction

As mentioned before, one of the biggest advantages of the MFM imaging is that it has allowed us to perform a large screening of parameters in Pt/Co/Ir multilayers that is our favorite system at CNRS to stabilize and observe isolated skyrmions at room temperature. To understand the influence of the multilayer parameters that are for example the thickness of the Co magnetic layer or the number of repetitions on the amplitude of the DM interaction and finally on the characteristics of the skyrmions, several series of multilayers grown by sputtering. Hereafter we present the results of this MFM study.

*Influence of the number of repetitions*

For this study, the trilayer is Ir1|Co0.6|Pt1 (thickness in nm) grown on a 10 nm-thick Pt buffer with different number of repetitions ranging from 1, 3, 5, 10, 20 to 30 times our trilayer. By AGFM and/or SQUID magnetometry, we found that the material parameters (magnetization at saturation, anisotropy energy, shape of the hysteresis loop) are changing from one sample to another (see Fig. 18(a) and (b)) but also that the magnetic configuration from the MFM images obtained on a 3× 3 µm$^2$ region are also showing some differences between the two multilayers with different numbers of repetitions. A first observation is that in the 20-times multilayer, the magnetic domains at remanence after demagnetization seem to be more ordered than the 5-times sample. In order to get

an estimation of the DM interaction, we have done the Fourier transformation on both MFM images and found that the mean domain periodicity for the 5-times sample is to be 206 ± 20nm and for the 20-times: 140 ± 20 nm, confirming that when increasing the number of repetition of the main trilayer, some of the intrinsic magnetic parameters of the system are changing. For these MFM images taken after demagnetization, we can use, as we developed for the STXM experiments, the mean domain width at remanence and compared to simulations. The best agreement is obtained for the two multilayers with 5 or 20 repetitions for |D| = 1.9 mJ.m$^{-2}$. This value is in quite good agreement with the value found in the system studied by STXM in section II.

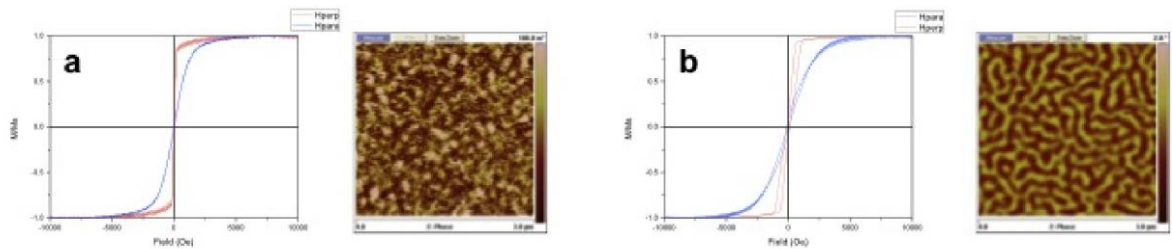

Figure 18: (a) Hysteresis loop and 3×3 µm² MFM image after demagnetization and at remanence of Pt10|{Ir1|Co0.6|Pt1}x3/Pt3 and (b) Hysteresis loop and 3×3 µm² MFM image after demagnetization and at remanence of Pt10|{Ir1|Co0.6|Pt1}x20/Pt3.

Beyond the estimation of DMi in the different multilayers, we have been also able to vary the perpendicular field in order to observe some magnetic skyrmions by MFM. Indeed, as shown in Fig. 19, we clearly observe some different magnetic patterns in the two multilayers with different repetitions. Indeed, both samples have been prepared with the same magnetic process, i.e. after demagnetization, we applied a large negative field (which is −150 mT with our MFM) and then slowly increasing the field towards positive values. Interestingly we notice the appearance of skyrmions do not show at the same field in the two multilayers and their density changes also. For the 5-repetition multilayer, we observe that the magnetic configuration remains saturated even at $\mu_0 H_\perp$ = −16 mT, meanwhile at $\mu_0 H_\perp$ = −18 mT on the 10-repetion one, we already observe a high density of skyrmions (see Figure 19b)). We also note that while the skyrmions are sufficiently separated from each other, their diameter vs $H_\perp$ = increases when the field becomes closer to zero as we observed by STXM on the same 20-repetition Pt/Co/Ir multilayer. However, it is interesting to see that, in the low field region, the density of skyrmions is increasing a lot (as the DMi amplitude is close enough to the threshold value for the onset of a skyrmion lattice phase) and consequently their diameters are decreasing due to the presence of the neighbors as it is observed in B20 skyrmion lattice phase.

From the evolution with $H_\perp$ = of the diameters of individual skyrmion, we deduce a DMi amplitude and find a good agreement between the experimental data and the simulated ones for these Pt/Co/Ir systems with DMi amplitude of 2.1 mJ/m², i.e. in agreement with the ones estimated from the domain periodicity. We also would like to emphasize that these MFM images are the first MFM observation of individual skyrmions in asymmetric multilayers at room temperature.

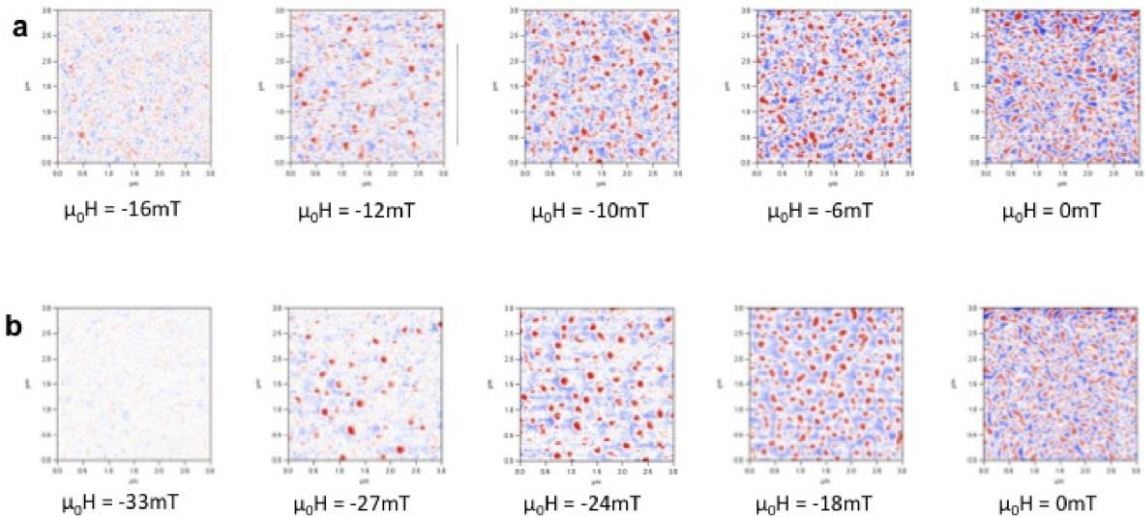

Figure 19: Room temperature MFM images (3×3 μm²) obtained (a) Pt10|{Ir1|Co0.6|Pt1}x5/Pt3 and (b) Pt10|{Ir1|Co0.6|Pt1}x10/Pt3 under different applied magnetic fields. Before imaging a relatively large negative fiel dis applied and then the field is slowly increased to enable the nucleation of magnetic skyrmions. We can notice here that the field value for which skyrmions are stabilized two multilayers are responding quite differently to out-of-plane magnetic field.

*Influence of the Co thickness in the multilayered stackings*

Our strategy to improve the thermal stability of skyrmions and more specifically to succeed to observe skyrmions at room temperature has been to increase the number of multilayers but at the cost of an increases of the demag pre-factor. Another way that we have studied by MFM is to increase the magnetic volume of our system was to increase the number of repetition of the main trilayer, and we have seen in the above paragraph that it might also influence the DM amplitude. Another way is to increase the individual thickness of the Cobalt layer. In the previous studies, we were estimating DMi with a Co thickness of 0.6 nm. The purpose of this paragraph is to highlight the impact of a thicker Co layer on the amplitude of DMi. Because the DMi in our system is interfacial, we would expect that its magnitude should decrease with thicker ferromagnetic layer.

For this study, three different samples with three different thickness to verify this theoretical prediction: Pt10|{Ir1|Co0.6|Pt1}x5/Pt3, Pt10|{Ir1|Co0.8|Pt1}x5/Pt3 and Pt10|{Ir1|Co1|Pt1}x5/Pt3 have been grown. A first result is that the multilayer is no more perpendicularly magnetized when the thickness of Co is higher than 1 nm or less than 0.6 nm.

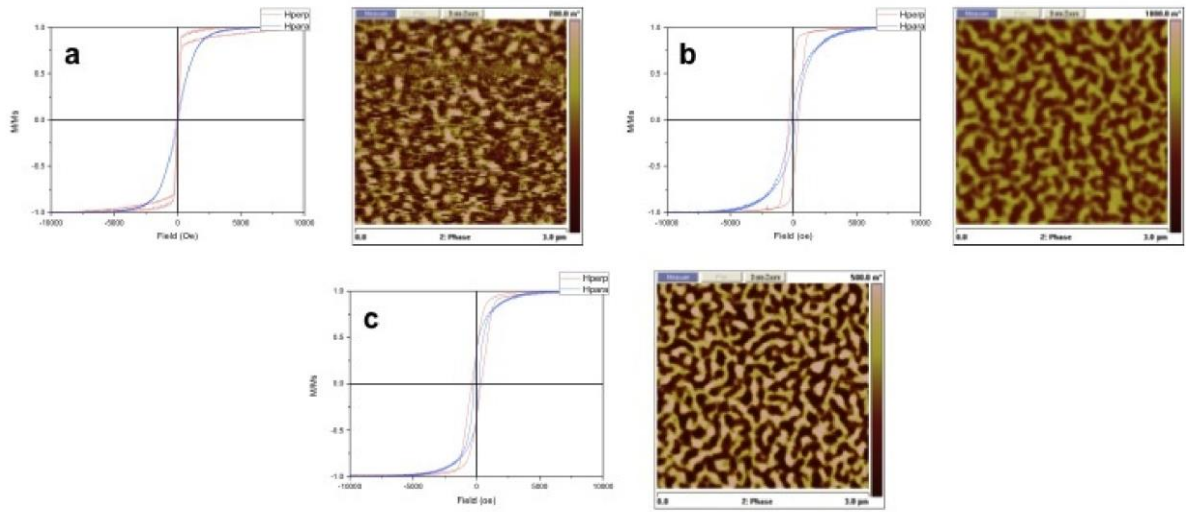

Figure 20: Hysteresis loop and room temperature MFM image (3×3µm²) after demagnetization and at remanence of Pt10|{Ir1|Co0.6|Pt1}x5/Pt3, (b) Hysteresis loop and MFM image after demagnetization and at remanence of Pt10|{Ir1|Co0.8|Pt1}x5/Pt3 and (b) Hysteresis loop and MFM image after demagnetization and at remanence of Pt10|{Ir1|Co1|Pt1}x5/Pt3.

Contrary to the case of multilayers with different number of repetition, the main magnetic parameters measured by magnetometry are not changing significantly for different Co thickness. However as shown in Fig. 20, we observe that the MFM images taken at remanence clearly show different patterns. As observed in the previous paragraph, we clearly observe a better contrast when increasing the thickness of Co, as expected. But, when considering the mean domain periodicity, we find : 206, 131 and 125 ± 20 nm for 0.6, 0.8 and 1 nm of Co respectively. Hence, there is a significant change between 0.6 and 0.8 nm of Co based samples. The change is not as important between 0.8 and 1 nm but the trend is similar : its seems that the thicker the Co, the smaller the mean domain periodicity. To get a more quantitative estimation of the DM interaction, we have performed the same analysis based on the domain periodicity after demagnetization or the evolution of the skyrmion size with $H_\perp =$. We deduce that DMi amplitudes decrease with thicker Co layers. Moreover, the difference between the 0.6 and 0.8 systems is bigger than between the 0.8 and 1 samples. Nonetheless, we have to keep in mind that the error margins for this estimation is about 0.2 mJ/m2.

### IV.3 Room temperature MFM observation of skyrmions in different systems of asymmetric multilayers.

We have also used the MFM technique for imaging skyrmions not only in extended films as presented before but in different types of nanostructures that have been designed and fabricated by standard electron beam lithography and ion beam etching process. These developments are part of our work program toward the elaboration of skyrmion based devices.

In this part, we present a collection of room temperature MFM images obtained on different multilayer systems.

*Tracks made from Pt/Co/Ir multilayers:*

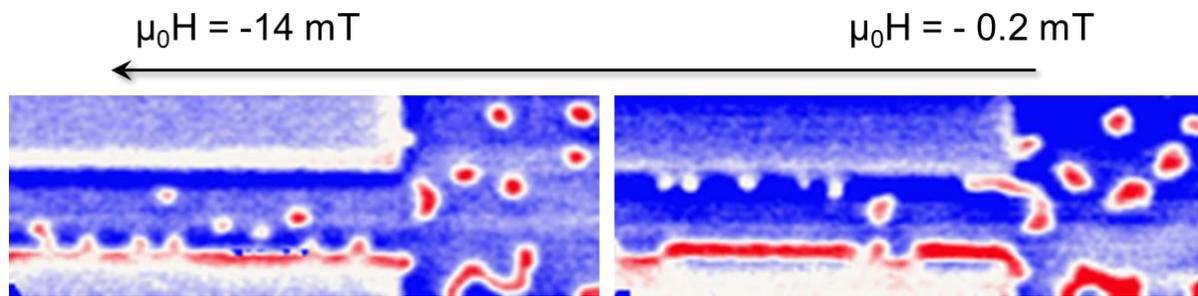

**Figure 21: Room temperature MFM image after demagnetization in a 700 nm wide track fabricated in Ta/Pt/[Co0.8/Ir1/Pt1]x20 multilayers.**

In Fig. 21, a series of MFM recorded at different applied fields in a 700 nm wide track prepared in a 20-repetition multilayer of Pt/Co/Ir. These devices are composed of a reservoir (on the right) connected to the strip. By sweeping the field, we can follow the evolution of the skyrmion diameter in the reservoir. Note that we can also image in the track the so-called "edge instabilities" (see J. Mueller et al, New Journal of Physics, 18, 2016) that are field polarized magnon states arising from the edges and that have been produced as a way to create skyrmions in nanostructures.

*Tracks and dots made from Pt/Co/W multilayers:*

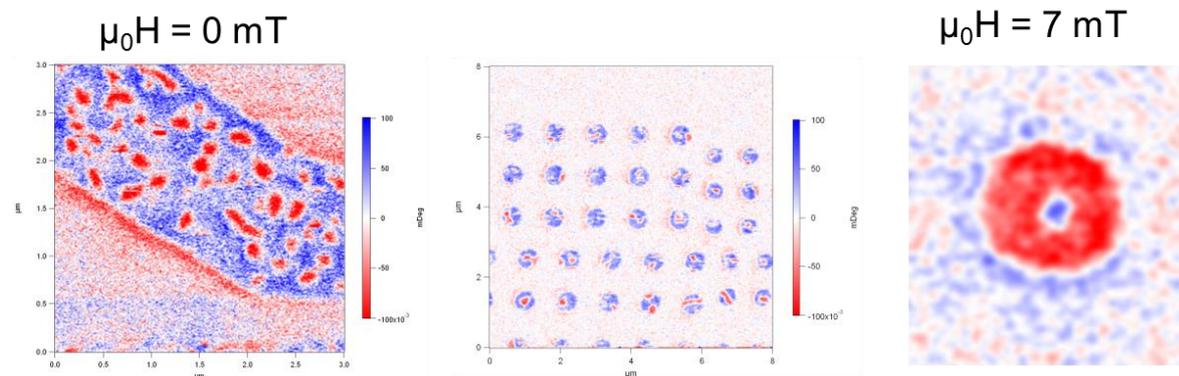

**Figure 22: (Left) Room temperature MFM image at zero field in a 2 μm wide track fabricated in Pt/{W1/Co0.6/Pt1}x10 multilayers (Middle) MFM images in an array of dots made from the same multilayers at very small perpendicular field; (Right) Isolated skyrmion stabilized at room temperature in a 600 nm diameter dot.**

In Fig. 22, we show a series of room temperature MFM images obtained in different types of nanostructures made from another asymmetric multilayer system, i.e. Pt/{W1/Co0.6/Pt1}x10. From the analysis of the domain periodicity in extended films (not shown) after demagnetization and the evolution of skyrmion diameter with perpendicular field (as presented before), we have been able to estimate that the DMi amplitude in this system is about 1.7 mJ/m² that is lower than the one from the Pt/Co/Ir system. However, this DMi value is large enough to allow the stabilization of skyrmions at room temperature. Interestingly, we see in Fig. 22(Left) that even at zero field, not all the skyrmions are transformed into elongated domains as we use to see in extended films, showing the role of the confinement and/or magnetostatic interaction for reinforcing the stabilization of skyrmions at room temperature. In the same multilayer system, arrays of sub-micrometer dots have been prepared in which isolated skyrmions can be imaged (see Fig. 22 (Right)), multilayers (Middle) MFM images in an array of dots made from the same multilayers at very small perpendicular field; (Right) Isolated skyrmion stabilized at room temperature in a 600 nm diameter dot.

*Tracks and made from Pt/Co/AlOx multilayers:*

Another multilayer system that we have investigated using the MFM technique is Pt/Co/AlOx. This system has been extensively studied by the Spintec and Néel groups in Grenoble (M. Miron et al, put refs) for the investigation of fast domain wall (DW) motion induced by spin orbit torques. The efficient DW dynamics is due to the fact that Néel DW are stabilized because of a large interfacial DM interaction of the order of 2.5 mJ/m². To our knowledge, this is the largest values measured experimentally both by DW nucleation analysis and by Brillouin Light Scattering (BLS).

We have grown a series of Pt/Co/AlOx multilayers (and not trilayers) in our sputtering systems and prepared some nano-tracks from them adapted for the injection of large current densities and electrical detection (see Figure 23) that will be investigated also within the MAGicSky project. As we presented for the other multilayer systems, from the analysis of the domain periodicity in extended films (not shown) after demagnetization and the evolution of skyrmion diameter with perpendicular field, we have been able to estimate that the DMi amplitude in our Pt/Co/AlOx is about 2.4 mJ/m² that is in agreement with values from the literature in this system.

In Fig. 23, we display a MFM image at room temperature obtained from sample composed of a reservoir on the left, the central track of 500 nm width and two lateral electrodes. This track has been prepared by e-beam lithography from Pt/{AlOx1/Co0.6/Pt1}x5. By applying a perpendicular magnetic field, we demonstrate that is possible to stabilize a discrete number of magnetic skyrmions arranged as a chain in the middle of the track.

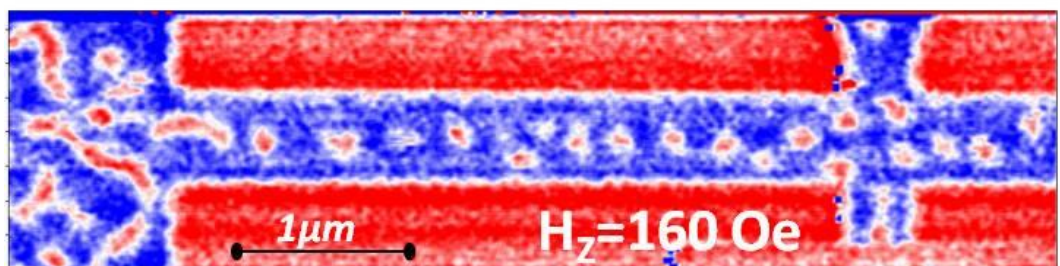

**Figure 23 : MFM image at RT in a 500 nm width Hall device made from Pt/{AlOx1/Co0.6/Pt1}x5. By applying magnetic field along the z-direction is possible to stabilize a discrete number of magnetic skyrmion arranged as a chain in the middle of the track.**

### IV.4 Summary and perspectives

In this section, we demonstrate that the MFM is a very efficient lab technique to provide some images of small skyrmions in multilayer-based systems at room temperature. Here we have used the MFM either to estimate the DM amplitude in extended films of different types of multilayers or to image isolated skyrmions in tracks or dots. As a perspective, we aim to simultaneously inject large current pulses for skyrmion displacement and image them by MFM, thus proving the principles of skyrmion-based racetrack.

# V/ State-of-the-art comparison worldwide

The recent demonstration of room temperature nanoscale chiral magnetic skyrmions led to an explosion of research interest in the international skyrmions community. Here we outline these international efforts by highlighting selected publications that are expected to define the field. We also detail how these research results and methods relate to the consortium results in the above sections. Specifically, we highlight the imaging methods used and their typical resolutions as well as the typical skyrmion sizes and the systems they were observed in.

**1)** *Real-space observation of a two-dimensional skyrmion crystal*, X. Z. Yu, Y. Onose, N. Kanazawa, J. H. Park, J. H. Han, Y. Matsui, N. Nagaosa & Y. Tokura, Nature 465, 901 (2010).

DMi can arise both from the breaking of the inversion symmetry at interfaces (interfacial DMi) or in the crystalline lattice itself (bulk DMi). B20 compounds such as $Fe_{0.5}Co_{0.5}Si$ exhibit such inversion symmetry breaking and are reported to exhibit skyrmion crystals under certain conditions, and, in fact, a lot of the fundamental work on skyrmions originates in bulk systems. Hexagonal skyrmion lattices in thin films of single crystalline $Fe_{0.5}Co_{0.5}Si$ (FZ grown and thinned down to 20 nm thickness by Ar-ion thinning) were imaged at temperatures below 30 K under applied out-of-plane magnetic fields between 30 mT and 80 mT with **Lorentz Transmission Electron Microscopy (Lorentz TEM)** (which exhibits a spatial resolution typically below 10 nm). The reported skyrmion lattice parameter was 90 nm. Our approach is to employ Lorentz TEM in microfabricated multilayer nanostructures that are more technologically relevant (ease of fabrication, tunability, room temperature) in relation to bulk samples.

**2)** *Room temperature skyrmion ground state stabilized through interlayer exchange coupling*, G. Chen, A. Mascaraque, A.T. N'Diaye, A. K. Schmid, Appl. Phys. Lett. 106, 242404 (2015).

G. Chen *et al.* in an early pioneering work describe the use of **spin-polarized LEEM (SPLEEM) imaging** for the investigation of the skyrmion ground state at room temperature of continuous films of Ni(15 ML)/Cu(x)/Ni(2 ML)/Fe(2-3 ML) (with different thicknesses of the Cu spacer layer) multilayers grown on Cu(001) substrates. Here, multiple skyrmions with a submicrometer reported size (~ 200 nm) were stabilized (at no applied external out-of-plane magnetic field). A spatial resolution of about 20-30 nm can be inferred from the reported SPLEEM images. Note here that the in-plane components of the skyrmion domain walls could also be completely resolved by the SPLEEM imaging.

**3)** *Room-temperature chiral magnetic skyrmions in ultrathin magnetic nanostructures*, O. Boulle, J. Vogel, H. Yang, S. Pizzini, D. de S. Chaves, A. Locatelli, T. O. Mentes, A. Sala, L. D. Buda-Prejbeanu, O. Klein, M. Belmeguenai, Y. Roussigné, A. Stashkevich, S. M. Chérif, L. Aballe, M. Foerster, M. Chshiev, S. Auffret, I. M. Miron & G. Gaudin, Nature Nanotechnology 11, 449 (2016).

In these experiments, O. Boulle *et al.* investigated micrometer-sized structures (fabricated by electron-beam lithography followed by ion milling) with different geometries out of thin films of Ta(3nm)/Pt(3nm)/Co(0.5-1nm)/MgO$_x$/Ta(1nm) multilayers exhibiting interfacial DMi with **XMCD-PEEM imaging** at room temperature. The group uses X-ray imaging similar to our results detailed in Section II and C. Moreau-Luchaire *et al.*, Nanotechnology 11, 444–448 (2016) but they use the Photoemission electron microscopy technique. We have focused on Scanning X-ray transmission microscopy because, in principle, it could give us excellent spatial resolution and superior temporal resolution in dynamics (sub-nanosecond) (see details in Section II). The stabilization of isolated skyrmions with a diameter between 70 nm and 190 nm (dependent on the magnitude of the externally applied out-of-plane magnetic field) was reported. Spatial resolutions down to 25 nm were reported, in ideal conditions, for XMCD-PEEM imaging. In the work presented in Ref. [Bou16], additional image processing was carried out, allowing for the identification of the size of the skyrmion domain wall with an error of 4 nm (at 95% confidence) as well as of the internal structure of the domain wall inside the skyrmionic spin texture they observe.

**4)** *Observation of room-temperature magnetic skyrmions and their current-driven dynamics in ultrathin metallic ferromagnets,* S. Woo, K. Litzius, B. Krüger, Mi-Young Im, L. Caretta, K. Richter, M. Mann, A. Krone, R. M. Reeve, M. Weigand, P. Agrawal, I. Lemesh, M.-A. Mawass, P. Fischer, M. Kläui & G. S. D. Beach, Nature Materials 15, 501 (2016).

In this work of S. Woo *et al.*, investigated multilayered {Pt(3nm)/Co(0.9nm)/Ta(4nm)}$_{15}$ and {Pt(4.5nm)/CoFeB(0.7nm)/MgO(1.4nm)}$_{15}$ exhibiting interfacial DMi with **XMCD-STXM** and **MTXM imaging** (with reported spatial resolutions of about 25 nm). Here, the groups use X-ray imaging similar to our approach detailed in Section II and C. Moreau-Luchaire *et al.*, Nanotechnology 11, 444–448 (2016). This technique has superior advantages in imaging not only statically but also in imaging the dynamics of magnetic domains (here, skyrmions). The films were patterned in micrometer (μm) and sub-μm structures with electron-beam lithography followed by lift-off. Both isolated magnetic skyrmions and hexagonal magnetic skyrmion lattices were observed for these multilayered materials at room temperature. Skyrmion diameters between < 50 nm and 250 nm (dependent on the magnitude of the externally applied out-of-plane magnetic field) were reported.

**5)** *Blowing magnetic skyrmion* bubbles, W. Jiang, P. Upadhyaya, W. Zhang, G. Yu, M. B. Jungfleisch, F. Y. Fradin, J. E. Pearson, Y. Tserkovnyak, K. L. Wang, O. Heinonen, S. G. E. te Velthuis, A. Hoffmann, Science 349, 283–286 (2015).

Magneto-optical imaging techniques, such as **Polar Kerr microscopy**, can also be employed for the investigation of skyrmions, as reported in this publication. However, due to the relatively low spatial resolution achievable by optical microscopy (diffraction limited), this technique is limited to the analysis of micrometer-sized skyrmionic bubbles. While this is highly interesting work, our main focus so far has been to initiate X-ray dynamical imaging so that we can access/image the dynamics technologically relevant skyrmions in the nanoscale and sub-nanosecond regime. In the work, microstructures of Ta(5nm)/CoFeB(1.1 nm)/TaO$_x$(3nm) multilayers were analyzed by dynamical Kerr microscopy (with temporal resolutions on the $10^{-1}$ s timescale). The stabilization of multiple skyrmions (not organized in a lattice) of diameters varying between 700 nm and 2 μm (dependent on the magnitude of the externally applied out-of-plane magnetic field) was reported.

**6)** *Tunable Room Temperature Magnetic Skyrmions in Ir/Fe/Co/Pt Multilayers,* A. Soumyanarayanan, M. Raju, A.L. Gonzalez Oyarce, Anthony K.C. Tan, Mi-Young Im, A.P. Petrovic, Pin Ho, K.H. Khoo, M. Tran, C.K. Gan, F. Ernult, C. Panagopoulos, arXiv:1606.06034 (2106).

MFM is another technique suitable for the investigation of skyrmions in magnetic multilayered films, as reported in this publication. As seen in this D1.1 deliverable report, we have already completed a full imaging study of multilayers with MFM (Section IV, partner CNRS) tuning their properties by engineering number of layers and interfaces. In this work by the Singapore group, micro- and nanodots with diameters between 500 nm and 3 µm were patterned by electron-beam lithography followed by ion milling out of {Ir(1nm)/Fe(*x*)/Co(*y*)/Pt(1nm)}x20 (*x* between 0 and 0.6 nm, and *y* between 0.4 and 0.6 nm) multilayer stacks, and imaged by **MFM** (typical spatial resolutions of about 30-50 nm, strongly dependent on the lift height of the MFM tip) and **MTXM**. Both isolated skyrmions and hexagonal skyrmion lattices were observed at room temperature, with reported skyrmion diameters between 50 and 70 nm (dependent on the magnitude of the externally applied out-of-plane magnetic field). A key point materializing in the community is the tunability of the magnetic skyrmions.

## Conclusion

In deliverable 1.1, we completed the imaging of individual skyrmions of multilayer systems made by different techniques using a plethora of key state-of-the-art imaging techniques. We engineered large DMis and demonstrated skyrmions by imaging several systems, from ultrathin films to multilayers, in both extended magnetic films as well as patterned nanostructures, which are essential for devices. Specifically, we i) fabricated several multilayer systems for skyrmion stabilization and ii) successfully demonstrated the imaging of nanoscale magnetic chiral skyrmions at room temperature and systematically investigated their magnetic fields stability. We achieved this by iii) engineering and evaluating large DMis in our material systems. We have also initiated the investigation in the thermal stability of skyrmions by imaging their behaviour from low temperatures (100K) up to room temperature (300K). Imaging the static behaviour of magnetic skyrmionic structures as well as their magnetic and thermal stability is paramount for their subsequent technological exploitation that MAGicSky aims to achieve. The aforementioned research results obtained as part of MAGicSky were disseminated in five internationally leading publications. These also attracted numbers of invited talks in the major conferences for physics, materials science, imaging, spintronics and skyrmions in particular, for the MAGicSky partners. Demonstrating and imaging magnetic chiral skyrmions at room temperature and zero magnetic field, confined in nanostructures from technologically relevant multilayers where a large DMi was engineered, is an international experimental breakthrough. It also is a major milestone upon which the MAGicSky research proposal will evolve. Imaging magnetic skyrmions will stand as a stepping-stone to every other experimental effort detailed in our proposal that will lead to proof-of-principle technological devices and we conclude that we have systematically completed it.